\documentclass[Afour,sagev,times]{sagej}

\usepackage{nameref}
\usepackage{moreverb,url}

\usepackage[colorlinks,bookmarksopen,bookmarksnumbered,citecolor=red,urlcolor=red]{hyperref}
\usepackage{booktabs}

\usepackage{multirow}

\newcommand\new[1]{{\color{black}#1}}

\newcommand\BibTeX{{\rmfamily B\kern-.05em \textsc{i\kern-.025em b}\kern-.08em
T\kern-.1667em\lower.7ex\hbox{E}\kern-.125emX}}

\def\volumeyear{2019}

\begin{document}

\runninghead{Furmanova et al.}

\title{Taggle: Combining Overview and Details in Tabular Data Visualizations}

\author{Katarina~Furmanova$^{*}$\affilnum{1},
        Samuel~Gratzl$^{*}$\affilnum{2}, 
		Holger~Stitz\affilnum{3}, 
		Thomas~Zichner\affilnum{4}, 
		Miroslava~Jaresova\affilnum{5}, 
		Alexander~Lex\affilnum{6}, 
		and Marc~Streit\affilnum{3}}

\affiliation{\affilnum{1}Masaryk University, Brno, Czech Republic\\
\affilnum{2}datavisyn GmbH, Linz, Austria\\
\affilnum{3}Johannes Kepler University, Linz, Austria,\\
\affilnum{4}Boehringer Ingelheim RCV GmbH \& Co KG., Vienna, Austria\\
\affilnum{5}Czechitas, z.s., Prague, Czech Republic\\
\affilnum{6}University of Utah, Salt Lake City, UT, USA\\
\affilnum{*}These authors contributed equally to this work.}

\corrauth{Marc Streit, Johannes Kepler University Linz, 
Institute of Computer Graphics, Altenbergerstraße 69, 4020 Linz, Austria}
\email{marc.streit@jku.at}

\begin{abstract}
Most tabular data visualization techniques focus on overviews, yet many practical analysis tasks are concerned with investigating individual items of interest. At the same time, relating an item to the rest of a potentially large table is important. 
In this work we present Taggle, a tabular visualization technique for exploring and presenting large and complex tables. Taggle takes an item-centric, spreadsheet-like approach, visualizing each row in the source data individually using visual encodings for the cells. At the same time, Taggle introduces data-driven aggregation of data subsets. The aggregation strategy is complemented by interaction methods tailored to answer specific analysis questions, such as sorting based on multiple columns and rich data selection and filtering capabilities. We demonstrate Taggle using a case study conducted by a domain expert on complex genomics data analysis for the purpose of drug discovery.
\end{abstract}


\maketitle

\renewcommand{\thefootnote}{*\arabic{footnote}}

\begin{figure*}[t]
 \vspace{-1mm}
\centering
\includegraphics[width=\linewidth]{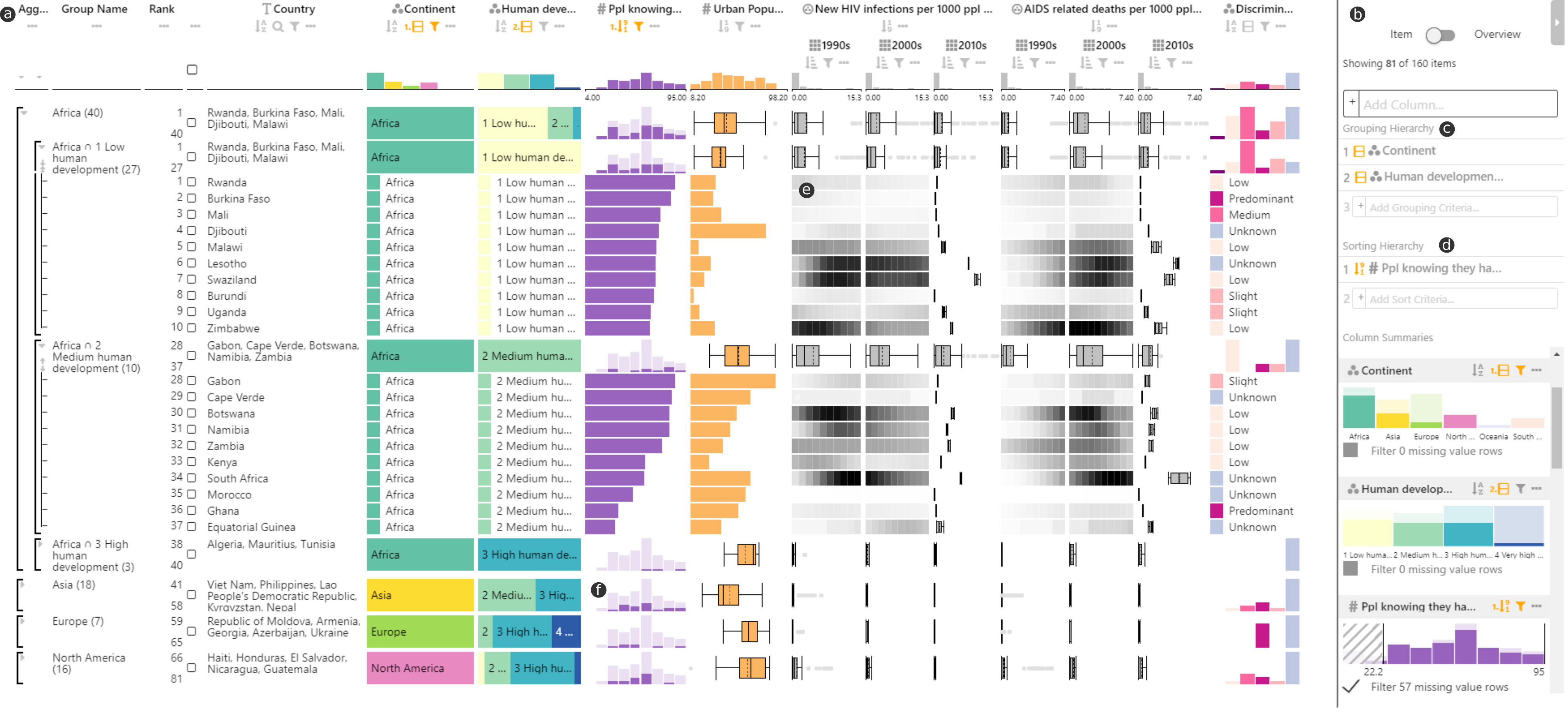}%[width=16cm]
 \vspace{-3mm}
\caption{The Taggle interface consisting of a table view (a) and a data selection panel (b) showing a dataset on AIDS in several countries grouped by continent and level of human development index. The data selection panel consists of grouping (c) and sorting (d) hierarchy panels and attribute filter views that allow users to filter out records by interacting with the histograms. The rows with individual African countries indicate the relationship between \textit{new HIV infections} and \textit{AIDS-related deaths} over time. It can be seen that an outburst of \textit{new HIV infections} in the 1990s in southern African countries resulted in high \textit{AIDS-related death} rates about a decade later in the 2000s (e). The rows of countries in \textit{Asia}, \textit{Europe}, and \textit{North America} have been aggregated to histograms, box plots, and stacked bars (f).}
 \vspace{-2mm}
\label{fig:aids}
\end{figure*}

\section{Introduction}
Visualization of tabular or multidimensional data is important in many application domains and is a mainstay of visualization research. Many multidimensional data visualization techniques, however, focus on providing overviews. To answer questions about the high-level similarity of items, projections techniques have proven useful, and exploring correlations between dimensions is well addressed by axes-based techniques such as scatterplot matrices and parallel coordinates plots. A third type of task is concerned with understanding the properties of an item in all dimensions, which is well addressed by tabular techniques. Tabular techniques use a spreadsheet-like layout, with each item in a row and each dimension in a column. In contrast to spreadsheets, the cells use visual encodings to make the data easy to view and to be able to explore higher level trends. Prominent examples of tabular visualization are the Table Lens~\cite{rao_table_1994}, Bertifier~\cite{perin_revisiting_2014}, LineUp~\cite{gratzl_lineup:_2013}, and ComplexHeatmap~\cite{gu_complex_2016}. 

A shortcoming of current tabular visualization techniques is their lack of sophisticated focus and context. A common solution implemented in both the Table Lens and LineUp is to scale down the rows in the visualization, and then use geometric distortion (lenses) to reveal details about selected items. Distortion, however, is associated with a variety of drawbacks, such as maintaining object constancy~\cite{munzner_visualization_2014}$^{,p.334}$. Also, lens-based approaches in tables rely on linear orderings, which cannot leverage higher level semantics of the data to provide compact but meaningful aggregations. Aggregation approaches based on grouping, in contrast, can stratify a table in a data-driven and hence semantically meaningful way. 

%In this paper, we introduce a tabular visualization technique that provides such data-driven aggregations. 
Our primary contribution is \textit{Taggle}, a tabular visualization method that displays large tabular datasets with up to a million data items by selectively grouping and aggregating subsets of a dataset. The goal of Taggle is to provide a high-level overview of large tabular datasets while allowing users to drill down to individual items. Groupings and aggregations of rows can be dynamically defined by users using selection, or in a data-driven way based on categorical or numerical dimensions. Hierarchical combinations of aggregations enable a fine-grained control of what to show in a dataset at which level of detail. Taggle also introduces grouping and aggregation of columns for cases where columns represent data of the same type, as, for example, in time-series data. The grouping and aggregation capabilities are complemented by sorting and filtering techniques. 

%When used with larger data sets, tabular data visualization typically requires a trade-off between overview and detail: for showing an overview, techniques such as the Table Lens down-scale rows to be as thin as a single pixel, whereas multiple pixels per row are necessary when labels are to be readable. However, even down-scaling rows has its limits, as information loss occurs when a dataset has more items than the screen has pixels.

%To this end, we use selective aggregation of rows and columns. Aggregation is enabled by various grouping strategies, which can be based on combinations of categorical attributes, user-driven selections, or on setting thresholds to numerical attributes. Both aggregates and individual rows can be visualized using a multiform approach, where users can choose the appropriate visualization techniques for the scale, aggregation status, and type of individual columns. 
%In this respect, Taggle fully generalizes our previous work on the LineUp~\cite{gratzl_lineup:_2013} visualization technique for multi-attribute ranking. 
%The resulting technique is suitable for both interactive exploration and presentation of complex tabular datasets. 
We showcase Taggle using a public health dataset: the spread of AIDS across the nations of the world. We also demonstrate Taggle using a variety of datasets, including a dataset of soccer players, programming language popularity, world happiness measures, economic data, and many others at~\url{https://taggle.caleydoapp.org/}. 

We \new{demonstrate} Taggle's utility by means of a case study on analyzing a cancer genomics dataset for the purpose of drug discovery.

%%%%%%%%%%%%%%%%%%%%%%%%%%%%%%%%%%%%%%%%%%%%%%%%%%%%%%%%%%%%%%%%

\section{Tabular Data}
\label{sec:data}
Throughout this paper, we use an AIDS dataset from UNAIDS AIDSinfo\footnote{http://aidsinfo.unaids.org/} as a guiding example. This dataset was enriched with metadata about the countries, such as population, which we retrieved from the United Nations Population Division\footnote{http://www.un.org/en/development/desa/population/} and the yearly Human Development Report of the United Nations Development Programme\footnote{http://hdr.undp.org/}.
The combined dataset consists of 17 numerical columns (e.g., \textit{population}, \textit{sex before the age of 15} in percent), 4 categorical columns (e.g., \textit{continent}, \textit{human development index}), and 10 time-series matrices (e.g., \textit{AIDS-related deaths} or \textit{new HIV infections over a period of 27 years}) collected for 160 countries.

Tabular datasets are usually composed of items stored in rows, which often correspond to independent variables (countries, in our example), and values (i.e., observations about these variables) stored in columns, which commonly correspond to dependent variables~(e.g., population or continent, in our example). Lex et al.~\cite{lex_visbricks:_2011} discuss heterogeneity and sources of heterogeneity in tabular data: \textbf{semantics}---the columns in the table have different meanings; \textbf{characteristics}---the columns have different data types and value ranges; and \textbf{statistics}---the columns have different behaviors or distributions. 

Homogeneous datasets lend themselves to compact and simple visual representations, as all data items share the same meaning and scales. Heatmaps~\cite{eisen_cluster_1998}, for example, are well suited to homogeneous datasets, as they encode each cell with a color value, which makes it possible to represent individual items at minimal scale.

Heterogeneous datasets have different semantics, characteristics, and statistics. Consequently, they may need separate scales and visual representations for each column. For instance, the~\textit{population} is given in absolute numbers and \textit{sex before the age of 15} is stated in percent.

We distinguish between the following data types: \textbf{Attribute columns} where all associated records are of the same type and semantics, such as the \textit{name}, \textit{gender}, and \textit{age} columns in a table of people. Attributes can be categorical, numerical, temporal (date and time), or textual. \textbf{Matrices} are composed of attribute columns of the same semantics and data type as is commonly found in, but not exclusive to, time series. An example is a country's \textit{GDP} over multiple years, where each year is a column in the matrix. A non-time-series example, common in the field of genomics, is a gene expression dataset, where the rows are genes and each patient is a column in the matrix. Although it is possible to interpret matrices as a list of columns, it is beneficial to treat them as a matrix, because the homogeneity of the data is an opportunity for compact representation. The columns in matrices can also be associated with attributes that describe a common property of the column, such as the decade associated with a year, or a shared phenotype of patients.

%%%%%%%%%%%%%%%%%%%%%%%%%%%%%%%%%%%%%%%%%%%%%%%%%%%%%%%%%%%%%%%%

\section{Design Goals}
\label{sec:design_guidelines}

Based on discussions with experts from various application domains who regularly analyze large tabular datasets, literature reviews, and our own experience, we developed a set of design goals for Taggle. Our first goal is to develop an \textbf{item-centric visualization technique that also explicitly shows all dimensions relevant to an analysis task}. This goal by itself is addressed by prior tabular data visualization technique, but currently no tabular data visualization technique addresses our second goal: \textbf{providing a seamless combination of overview and details through selective, data-driven aggregation}. A technique that would satisfy this goal would remedy the major drawback of tabular data visualization techniques: limited context. Current tabular visualization techniques can only provide context only by showing neighbors through a single, global sorting, which makes it difficult to compare items of different categories. This design goal is hence concerned with showing the details about selected items and providing context, e.g., through aggregations of data-driven groups. 

To fully leverage the potential of an overview plus detail tabular data visualization technique, we need to give users the ability to flexibly define the parameters of the display. To address that, our third goal is to provide \textbf{rich interaction techniques that support answering specific questions, such as sorting, filtering, and grouping}. Finally, to appropriately visualize the diverse data types and different levels of aggregations, we need to \textbf{provide a variety of visual encodings suitable for the specific situations}. One goal is to provide sensible defaults, but we also need to provide the ability to \textbf{flexibly choose visual encodings tailored to data types and aggregation levels}, to account for the diversity of analysis questions and dataset characteristics.

%%%%%%%%%%%%%%%%%%%%%%%%%%%%%%%%%%%%%%%%%%%%%%%%%%%%%%%%%%%%%%%%

\section{Related Work}
\label{sec:related_work}

We discuss related work in light of two considerations: (1)~a review of tabular data visualization techniques, and (2)~approaches to aggregation.

%\todo{, grammel_survey_nodate, jo_touchpivot:_2017}, possibly also \cite{grammel_how_2010}} ? \cite{jentner_concept_2018}

\subsection{Tabular Data Visualization}

Since tabular data analysis plays an important role in many research fields, a substantial body of work exists on visualizing such data. We distinguish between four types of tabular data visualization techniques: 
\begin{enumerate}
  \item \textbf{dimensionality reduction techniques}, which show a lower dimensional projection of a high-dimensional dataset,
  \item \textbf{axes-based techniques}, which position marks for each cell based on its value, such as parallel coordinates, star plots, and scatterplot matrices,
  \item \textbf{tabular techniques}, which retain item positions across columns and encode the data within the cells,
  \item \textbf{multiple coordinated view (MCV) and hybrid techniques}, which show visualization of individual dimensions or subsets of attributes in separate but linked views. 
\end{enumerate}

\new{Our four types of tabular data visualization techniques are related to the three families proposed by Dimara et al.~\cite{dimara_conceptual_2018}. In their work, they distinguish between lossy and lossless geometric projection techniques. Lossy techniques do not preserve the raw values of individual dimensions, i.e., this category corresponds to the dimensionality reduction techniques. Their family of lossless techniques includes axes-based and tabular techniques, which we keep separate, as they employ different data encoding principles.}

\subsubsection{Dimensionality Reduction Techniques}

Projection or dimensionality reduction techniques techniques visualize the structure of items associated with high-dimensional data in a lower dimensional space. There are various commonly used approaches, such as principal component analysis, multidimensional scaling techniques, or t-SNE~\cite{vandermaaten_visualizing_2008}. For data visualization, usually a 2D or sometimes also a 3D representation of the projected items is displayed. These low-dimensional projections show groups of similar items close to each other. One problem of projections is that they can produce artifacts showing items that are quite different in proximity. A variety of techniques have been proposed to address this and related shortcomings~\cite{chuang_interpretation_2012, stahnke_probing_2016, sacha_visual_2017}. Another challenge with dimensionality reduction is the sensitivity of the results to the choice of algorithm and the sensitivity to parameters~\cite{wattenberg_how_2016}, which often makes an iterative approach with multiple parameters and/or algorithms necessary. 

\new{A special case of dimensionality reduction is to turn relationships and items into a network, and then render that network using, for example, force-directed layout algorithms. Examples of this approach are Ploceus~\cite{liu_ploceus:_2014}, Orion~\cite{heer_orion:_2014}, and Origraph~\cite{bigelow_origraph:_2019}.}

We argue that projection techniques are well suited to visualize structure in a high-dimensional dataset, but they cannot adequately show \textit{why} items in a cluster belong together. Projection techniques are especially useful in cases where the dimensions themselves are not meaningful to human analysts, such as a table of term frequencies when analyzing text documents. Taggle is concerned with exactly the opposite use cases: where the properties of the dimensions are critical in making decisions.

\subsubsection{Axes-based Techniques}

Axes-based technique use axes representing individual attributes and spatially encode the items' values. Key examples are scatterplot matrices~\cite{becker_brushing_1987, elmqvist_rolling_2008}, which place scatterplots consisting of orthogonal axes to show pairwise relationships between attributes in a matrix, and parallel coordinates~\cite{inselberg_parallel_2009, wegman_hyperdimensional_1990, heinrich_state_2013}, which place axes in parallel and connect individual items to their position on the axes using polylines. 
Variations of parallel coordinates are star plots~\cite{kandogan_visualizing_2001}, where all axes originate from a common point, or other, more general axes-based layouts~\cite{tominski_axes-based_2004}. Generalizations of axes-based techniques include FLINA~\cite{claessen_flexible_2011}, a technique that lets users flexibly arrange axes and choose between connection lines or dots, and GPLOM~\cite{im_gplom:_2013}, which generalizes the scatterplot matrix idea to other visualization techniques shown in the cells. 

Axes-based techniques can effectively show correlations between neighboring axes. However, the quality of insights depends on the order of the axes. Other limitations are the visual clutter caused by crossing polylines and the fact that axes-based technique are problematic for encoding categorical and textual attributes.

\subsubsection{Tabular Techniques}

Tabular visualization techniques use a grid layout where rows represent items and columns dimensions (although the inverse is also possible); the value of each item in each dimension is encoded in a cell. Within the class of tabular techniques, we further distinguish tabular visualizations for homogeneous tables, visualizations for heterogeneous tables, and spreadsheet tools. \new{An overview summarizing the features and supported tasks of individual tabular visualization techniques listed in this section can be found in Table 1 of the supplementary material.}

The prototypical example of a \textbf{homogeneous tabular visualization} technique is a heatmap~\cite{wilkinson_history_2009}, where cell values are encoded using color (hue, saturation, value, or opacity). Homogeneous table visualization tools are useful for data that has the same type and scale across all dimensions (matrices, according to our definition in Section~\nameref{sec:data}). Heatmaps are exceptionally scalable, as the cells can be allocated as little as a single pixel of space. A key aspect is to find good orderings of the rows and columns, which is often done using clustering~\cite{eisen_cluster_1998} or seriation approaches~\cite{liiv_seriation_2010}. Visualization tools that provide advanced features for heatmaps include the Hierarchical Cluster Explorer~\cite{seo_interactively_2002}, GAP~\cite{wu_gap:_2010}, PermutMatrix~\cite{caraux_permutmatrix:_2005}, Clustergrammer~\cite{fernandez_clustergrammer_2017}, and SmartExplore~\cite{blumenschein_smartexplore:_2018}. Taggle can efficiently visualize homogeneous tables, but in contrast to the techniques discussed here, Taggle also supports heterogeneous tables, and can combine homogeneous parts of a heterogeneous table (matrices) and heterogeneous columns in a single visualization. %Examples of tools that combine homogeneous and heteroegeneous tables are ComplexHeatmap~\cite{gu_complex_2016}, an R package that allows users to combine heatmaps for matrices with categorical and numerical vectors using flexible encoding possibilities, and the table view in EnRoute~\cite{partl_enroute:_2012}. 
% It uses various visual encodings for cells that can be interactively re-ordered based on similarities between rows and columns. Other features, such as styling options for the table grid, indicate that the technique is intended mainly for presenting small or medium-sized tables. Bertifier does not support aggregation or grouping and is limited to numerical data.

The Table Lens~\cite{rao_table_1994} is a \textbf{tabular visualization technique suitable for heterogeneous tables}. It is probably the most closely related technique to Taggle and inspired its development. It uses visual encodings tailored to different data types to represent values in cells. Rich sorting operations allow users to compare trends between separate attributes. Scalability is achieved by downscaling rows, and a combination of appropriately chosen visual encodings and lens techniques ensures readability of trends and individual items. The most important differences to Taggle are that the Table Lens does not support aggregation and is therefore limited in terms of scalability. Taggle also introduces a variety of subtle new ideas, such as embedding space-efficient techniques for homogeneous subsets of a table. A variety of tools, such as DataComb~\cite{polis_datacomb:_2015}, the Visual Spreadsheet~\cite{goldman_visual_2017}, and the table views in some multivariate tree and network visualization tools~\cite{nobre_lineage:_2019, nobre_juniper:_2019} implement ideas of the Table Lens. \new{Another technique employing various visual encodings suitable for heterogeneous tables is Bertifier~\cite{perin_revisiting_2014}. It was inspired by Jacques Bertin's matrix analysis methods and supports interactive data reordering based on similarities between rows and columns. However, the technique is intended mainly for presenting small- or medium-sized tables.}

Widely used \textbf{spreadsheet tools}, such as Microsoft Excel\footnote{\label{excel}https://products.office.com/en-us/excel/},
Google Sheets\footnote{https://www.google.com/sheets/about/},
and Apache OpenOffice Calc\footnote{https://www.openoffice.org/product/calc.html}
typically support tabular operations such as sorting, filtering, and grouping. However, although spreadsheet tools usually support rich charting operations, they provide only limited support for direct visual encoding of cells, using techniques such as conditional formatting.
%\textit{Tableau}~\cite{stolte_polaris:_2008} can be used to create tabular visualizations that use a variety of visual encodings. However, while Tableau supports most standard operations, it requires the manipulation of a complex interface that must be learned.

FOCUS~\cite{spenke_focus:_1996} and its successor InfoZoom~\cite{spenke_infozoom-analysing_2000} are hybrid spreadsheet/tabular visualization tools. In addition to the Table-Lens-like layout, InfoZoom provides an overview mode that shows the distribution of values for individual attributes, sorting each attribute row individually. Although this provides an overview of the distribution of values, it is no longer a tabular layout.

\subsubsection{Multiple Coordinated View Techniques and Hybrids}

Multiple coordinated view (MCV) systems represent (sets of) attributes of a tabular dataset in separate, linked views. These systems allow users to choose representations that are suitable for the subset of data represented by a single view, and usually rely on linked highlighting to highlight the same items in different views. Representative systems in this category include Improvise~\cite{weaver_building_2004} and Keshif~\cite{yalcin_keshif:_2018}. Common configurations of Keshif, for example, use a tabular view to identify specific items, but represent other attributes in other views using histograms or bar charts, for instance.

Although MCV systems can leverage visualization techniques that are ideal for certain attributes and that would potentially not fit into the confines of a tabular layout, they also add complexity and increase the cognitive load for the user~\cite{baldonado_guidelines_2000}. Tabular layouts, in contrast, make the association of all attributes to their item easy, but make it harder to see correlations between attributes or trends across the whole dataset. 

As the Keshif example shows, tabular visualization techniques, such as Taggle, are an ideal complement to MCV systems: although selected attributes can be shown in dedicated views, for example, on a map or in a node-link layout, other attributes can be shown as part of the tabular visualization.

Note that the line between MCVs and other techniques is fluid; a scatterplot matrix, for example, can be considered as both an axes-based technique and an MCV system.

\textbf{Hybrid approaches} that use multiple views and combine overview and tabular approaches or overview and projection approaches are also available. In hybrid overview-tabular approaches, the rows are preserved within subsets of the data, but the relationships between subsets are visualized using an overview technique. Examples of this class include NodeTrix~\cite{henry_nodetrix:_2007}, VisBricks~\cite{lex_visbricks:_2011}, StratomeX~\cite{lex_stratomex:_2012a,kern_interactive_2017}, Domino~\cite{gratzl_domino:_2014}, and Furby~\cite{streit_furby:_2014}. In hybrid overview-projection approaches, selected attributes are plotted on top of a plot of projected data, as in the technique developed by Stahnke et al.~\cite{stahnke_probing_2016}. Domino~\cite{gratzl_domino:_2014} is a hybrid tabular/overview MCV technique. It is based on the concept of placing subsets of a dataset on a canvas and choosing a suitable representation (view) for it. Multiple subsets can then be connected to show their relationships in various ways. Matchmaker~\cite{lex_comparative_2010}, VisBricks~\cite{lex_visbricks:_2011}, and StratomeX~\cite{lex_stratomex:_2012a, kern_interactive_2017} are related hybrid techniques, but they are more restricted with respect to the selection and layout of subsets. 

\subsection{Aggregation Methods}

Orthogonal to the design space discussed above are aggregation methods for tabular data: representing the underlying distribution or statistical measures of a set of items is an important approach to increasing the scalability of visualization techniques. Aggregation can be applied to a whole dataset or to multiple groups of items and/or attributes separately. Elmqvist and Fekete~\cite{elmqvist_hierarchical_2010} proposed several design guidelines for aggregation, including: \textit{Visual Summary}---aggregates should convey information about the underlying data; \textit{Discriminability}---aggregates can easily be distinguished from individual data items; and \textit{Fidelity}---measures are taken to counteract artifacts of the aggregation process that misrepresent true effects. The aggregation techniques in Taggle were designed with these guidelines in mind.

Examples of \textbf{overview techniques} using aggregation are hierarchical parallel coordinates~\cite{fua_hierarchical_1999}, which visualize cluster centroids rather than individual items, and VisBricks~\cite{lex_visbricks:_2011}, which can visualize clusters using various techniques, including statistical summaries such as histograms. An example \textbf{MCV technique} that predominantly uses aggregations is Keshif~\cite{yalcin_keshif:_2018}. In Keshif, a table of items is supplemented with multiple views showing distributions for interaction-driven exploration. 

To our knowledge, there is currently no interactive \textbf{general tabular visualization technique} that allows aggregation. When working with large tabular data, not all data can be shown in detail, as the number of rows quickly exceeds the available display space. There are two potential remedies: scrolling and aggregation. Although scrolling is common when working with tables, it does not preserve the context of off-screen data items. Aggregation, in contrast, can be leveraged to preserve both details about a set of items in focus and context about the rest. 

Various specialized tabular visualization tools use aggregation in tabular layouts. iHAT~\cite{heinrich_ihat:_2012} aggregates amino acid sequences and associated metadata using the most frequent category or the average to represent aggregated items, depending on the data type. Holzh{\"u}ter et al.~\cite{holzhuter_visualizing_2012} use the average for numerical values for aggregates. Both techniques employ transparency to communicate fidelity (the higher the variation in a cell, the higher the transparency), but neither addresses fidelity well. 
The Breakdown Visualization technique by Conklin and North~\cite{conklin_multiple_2002} aggregates rows or columns of a table based on a pre-existing aggregation hierarchy. Users can traverse the hierarchy and pivot through intersecting hierarchies. 
The UpSet~\cite{lex_upset:_2014} technique aggregates items based on set memberships. It uses visualizations such as box plots for representing aggregated group statistics.
In contrast to these techniques, Taggle provides the user with the flexibility to aggregate subsets of the table, while keeping details of other parts of the table visible in place.

\section{Visualization and Interaction Design}
\label{sec:concept}

Taggle is an item-centric visualization technique that shows all dimensions relevant to an analysis task and at the same time provides a seamless combination of overview and details through selective, data-driven aggregation. Here we introduce this approach.

%We start this section by introducing the hierarchical grouping and aggregation mechanism of Taggle, which provides the conceptual foundation for all further operations, such as filtering, sorting, and flexible encoding of the different subsets. We then discuss the layout strategy that enables Focus+Context also for large tabular datasets.

%\subsection{Hierarchical Grouping and Aggregation}

\begin{figure}[t]
\centering
\includegraphics[width=\columnwidth]{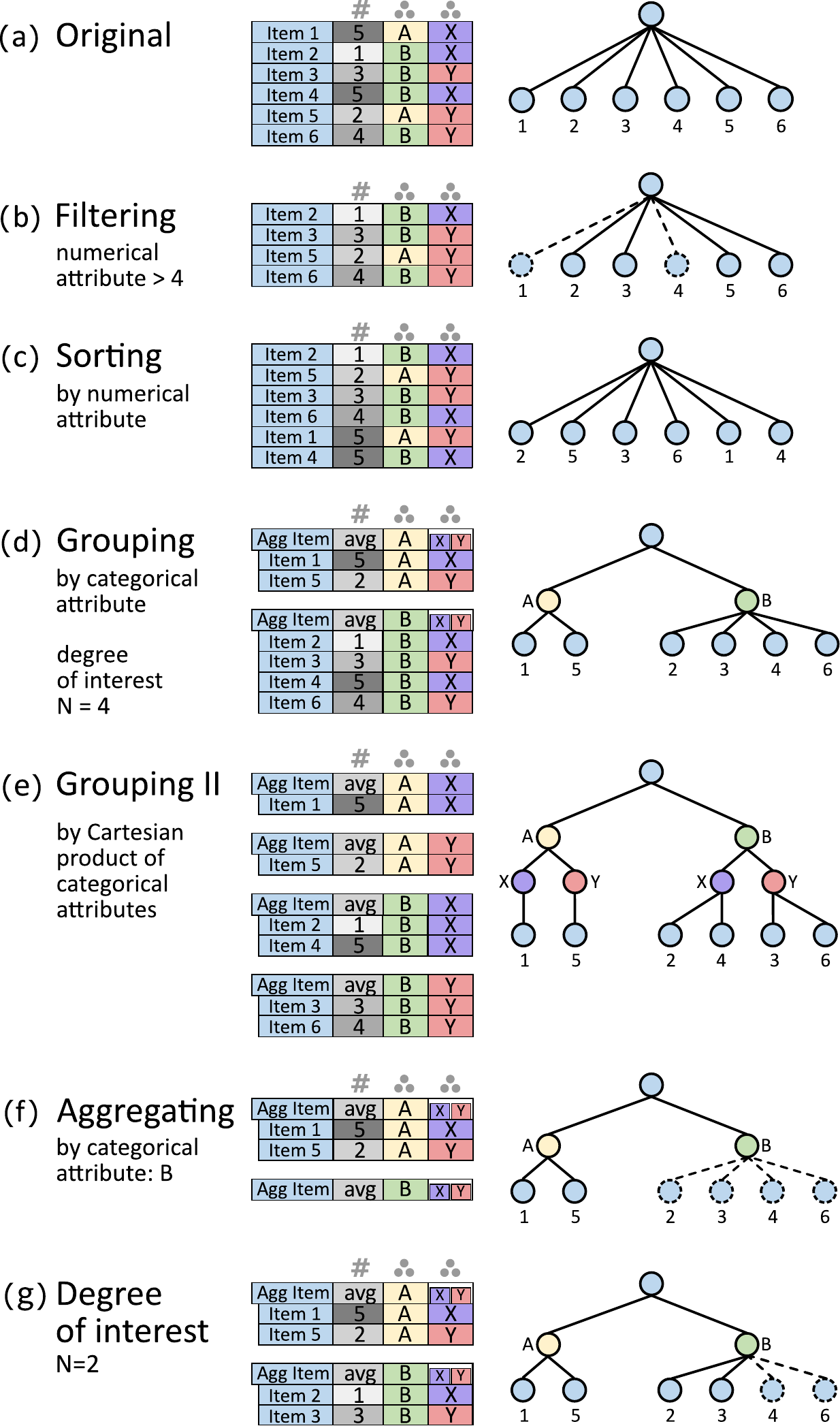}
\vspace{3pt}
    \caption{Illustration of topological operations on a heterogeneous table (a) consisting of numerical (\protect\includegraphics[height=0.7em]{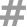}) and categorical (\protect\includegraphics[height=0.7em]{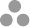}) attributes and their results reflected in the aggregation hierarchy: filtering (b), sorting (c), grouping by a single categorical attribute (d), grouping by the Cartesian product of two categorical attributes (e), aggregating (f), and degree of interest (g).
    }
    \label{fig:topological_operations}
\end{figure}

Taggle enables users to group items based on hierarchical combinations of attributes. The result of these nested grouping levels is an ordered tree where all leaves are items (Figure~\ref{fig:topological_operations}~(a)). Data-driven filter and sorting operations (Figure~\ref{fig:topological_operations}~(b) and (c)) can be used to reveal items of interest. 

By defining groups, we can add new levels to the tree (Figures~\ref{fig:topological_operations}~(d) and (e)). For example, we can group the countries in the AIDS dataset by continent. Groups can be defined based on categorical attributes, numerical thresholds, or user selections. Groups are represented as a row showing summary representations for the items in the group. 

Each branch in the tree can be collapsed independently, hiding the items while the group summary remains, as shown in Figure~\ref{fig:topological_operations}~(f). Each row of the resulting table then corresponds to either one item or one group. We can use this approach, for example, to show summaries of all continents, but also to show the individual countries on the African continent. By adjusting the level at which to aggregate, users can dynamically control the level of detail of the rows when rendering the table~\cite{elmqvist_hierarchical_2010}. 

Finally, we introduce a degree of interest operation~\cite{furnas_generalized_1986} to reveal aggregated items that are especially relevant to the analysis. \new{Our current implementation is naive, revealing only the first N items of an aggregated group. By leveraging sorting, we ensure that these items are the most relevant to the current analysis task. The operation allows us, for example, to show a summary about the AIDS epidemic by continent and reveal the ten most affected countries for each continent at the same time.
The degree of interest can be adjusted to reveal more or fewer items (Figure~\ref{fig:topological_operations}~(g)). This function could be improved to take other aspects of the data into account, such as a cut-off of an attribute or the size of the group.}

%Alternatively, each group can be expanded to show all of its items.

%Sorting, filtering, grouping, and aggregating are topological operations that result in changes in the tree. In contrast, encoding changes
%correspond to property changes in the nodes. Figure~\ref{fig:topological_operations} illustrates how the different topological operations alter the tree. By default, all rows of the table are added as nodes attached to a common root node. Filtering items from the table removes the corresponding nodes from the tree. Sorting of items or groups changes the order of nodes in the tree. Grouping items adds a level to the hierarchy. 

%To determine which rows to render, we use a level traversal strategy~\cite{elmqvist_hierarchical_2010}. In an un-aggregated tree, this traversal strategy results in a list of rows that correspond to the items of the table. When a node is aggregated, the traversal stops at this point, and adds only one row for this non-leaf node, as indicated by the stippled horizontal line shown in Figure~\ref{fig:topological_operations}~(f).

%Note that defining hierarchies and groups on the columns of a matrix works in the same way as for rows.

\subsection{Overall Design}

The Taggle interface consists of two parts, as shown in Figure~\ref{fig:aids}: (a) the main \textbf{table view} and (b) a \textbf{data selection panel} that is the interface for various operations.
The table view implements the overview plus detail concept for visualizing tabular data. The column headers of the table view provide the means for sorting, changing visual encoding, filtering, and grouping. %Users can select one or multiple items that will be highlighted by brushing over them while holding down the mouse or by checking a selection box offered for each item. In the overview, selected items will be increased to a predefined fixed height that allows labels to be shown.}
The data selection panel provides access to all available numerical, categorical, text, and matrix attributes. Its primary use is to enable analysts to choose which attributes to show in the table view. 
%\rem{When a vector or a matrix has been selected from a list of all loaded attributes, it is added as the last column to the table view. The user can change the order of columns via drag and drop.} 
For each column that is shown in the table view, the data selection panel shows a visual summary of the data in the form of a histogram, when appropriate. Below, we introduce the visual elements and interactions in detail, together with justifications of our design decisions.

\subsection{Layout Strategy}

%three detail levels (states):
%1. aggregated group
%three representations: 
%a aggregated group summary with fixed height
%b aggregated group summary plus N samples
%c aggregated group summary with all items
%2. space-filling group = flexible item height
%3. detail = fixed item height (always used for selected items)

% describe overview and detail mode as predefined configurations that use the atomic states from above. other configurations are possible as well (but not implemented)

Complementary to our overview plus detail concept described above, we introduce two different layout modes serving the high-level tasks of (1) obtaining an \textbf{overview} and (2) seeing \textbf{details} for a subset of the items. 

The goal of the \textbf{detail mode} is to allow users to see all details for selected items including labels, numerical values, and category names. Although this maximizes the readability of items, it comes at the cost of reducing the number of visible items. 

%While aggregation is a viable strategy to show information about all items, it relies on analysts being able to define meaningful groups. 

In \textbf{overview mode}, the goal is to show as many rows as possible in order to give users a good sense of the overall patterns and distributions. To achieve this, Taggle decreases the height of items until the whole table fits on the screen, or until each item has a height of a single pixel, as lower values would introduce uncertainty due to interpolation artifacts~\cite{holzhuter_visualizing_2012}. Aggregated groups are shown using a fixed height. Overview mode is a complementary strategy to aggregation: it is useful to get an idea about the distribution of the data in the columns and does not require that meaningful groups are defined. 
%Consequently, for tables that have more rows than can fit within the window, the table representation exceeds the available screen space and scroll bars are introduced. If this happens, the user has two options to make the table fit on the screen again: aggregating groups or filtering items.
When viewing the table in overview mode, users can still increase the level of detail for one or multiple items by selecting them, which is useful in cases where users spot items of interest that they want to inspect in detail.

\subsection{Sorting}
\label{ssec:sorting}

Sorting is a simple way of identifying minima and maxima in columns. Sorting also reveals relationships between columns. 
In addition to sorting in ascending or descending order by a numerical, textual, or categorical column, Taggle enables users to sort items hierarchically, where a top-level column determines the initial sorting, a second column breaks ties from the initial sorting, and so on. This sorting strategy is particularly useful when sorting by categorical columns. Users can also sort matrix columns by specifying a statistical measure (minimum, maximum, lower and upper quartile, median, mean) as the sorting criterion. 

Although other table visualizations such as the Visual Spreadsheet~\cite{tyner_ucsc_2017} sort attributes hierarchically based on the order of the columns, we decided to separate the sorting from the layout. Since we expect that in most cases users are satisfied with simple sorting by one attribute, clicking on the sort button in the column header always results in the data being sorted by the corresponding attribute. Once the user activates the sorting by one attribute, a dedicated sorting hierarchy panel appears in the data selection panel. The panel allows users to add additional sorting attributes and change their order (see Figure~\ref{fig:aids} (d)).

\subsection{Filtering}
Filters can be defined by interacting with the histograms in the data selection panel either by brushing a range in the case of numerical data (Figure~\ref{fig:aids} (b), \textit{people knowing they have HIV}) or by selecting categories that are to be removed from the table (Figure~\ref{fig:aids} (b), \textit{continent}). Textual data can be filtered by string matching or by a regular expression. In addition, users can filter out items with missing values. As an alternative to setting filters in the data selection panel, users can open a filter dialog via the header of the columns. 

\begin{figure*}[t]
\centering
\includegraphics[width=1\linewidth]{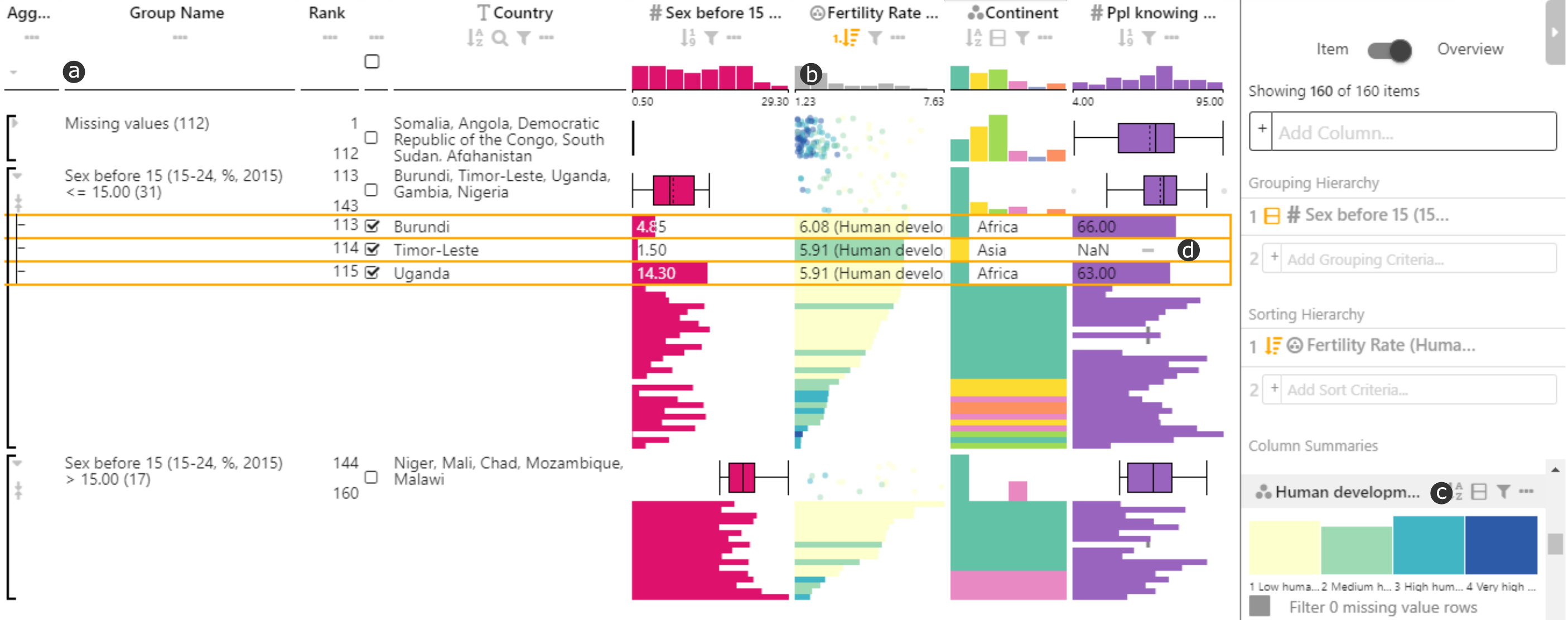}
\caption{Taggle table showing countries grouped by bins of the percentage of the population who had \textit{sex before the age of 15} (a). The \textit{fertility rate} values (b) are colored according to the \textit{human development index} (c), showing the correlation between the two attributes. Missing values are encoded using a dash (d).}
\label{fig:fertility}
\end{figure*}

\subsection{Grouping and Aggregation}
\label{ssec:grouping}

Being able to stratify tables into meaningful groups is not only an important feature for structuring tabular data, but also an essential prerequisite for aggregation operations in Taggle.

Grouping is related to sorting since grouping also influences the order of items. Taggle, however, separates these operations in order to enable more fine-grained control of groups. As discussed before, we leverage categorical or binned numerical attributes to group datasets. Similarly, we can leverage regular expressions on string columns to create groups, or use dates and date ranges on date columns. Users can also split the table into two groups based on the current selection. Combining multiple hierarchically sorted grouping attributes creates fine-grained groups that correspond to the Cartesian product of the constituting categories. In practice, we found that two to three grouping levels are sufficient, because more lead to fragmented groups.

Setting the grouping hierarchy is analogous to hierarchical sorting---the order of grouping attributes is indicated in a dedicated grouping hierarchy panel. Since grouping takes precedence over sorting, the hierarchy is shown above the sorting hierarchy panel (see Figure~\ref{fig:aids} (c)). The separation of grouping and sorting operations gives the user tighter control over the order of the table items. Users can, for example, group items based on a binned numerical attribute but sort the items inside the groups according to a different attribute.

A group name column summarizes the current grouping and how many items are contained. In Figure~\ref{fig:aids}, for instance, the combination of the attributes \textit{continent} and \textit{the human development index} constitutes the grouping, which is indicated in the first column. Groups can also be sorted by their name, by the number of contained items, by statistical measures of numerical attributes (e.g., mean or median), or by the most frequent category. Selected options are shown in an additional group sorting hierarchy in the panel.

Figure~\ref{fig:fertility} illustrates a case in which the countries were first grouped based on the percentage of women who had \textit{sex before the age of 15} with a threshold set to 15 percent (Figure~\ref{fig:fertility} (a)), but sorted according to \textit{fertility rate} (Figure~\ref{fig:fertility} (b)). Interestingly, only African and North American countries fell within the group with high percentages of \textit{sex before the age of 15}. Sorting the table by \textit{fertility rates} shows a clear difference between the countries of the two continents, with North American countries having much lower fertility rates than the African countries in this group. Fertility rate also correlates inversely with the level of \textit{human development index}. 

Groups are represented by rows showing an aggregate of the items they contain. Group headers are assigned a uniform height that is about twice that of a row shown in detail mode. We use dedicated visual encodings for aggregate items. For example, instead of bar plots for individual items, we show a histogram or a box plot that represents the whole group (see Figure~\ref{fig:aids} (f)). As discussed earlier, the items in a group can be shown below its header, partially hidden based on a degree of interest function, or completely hidden. 
%User can toggle the aggregation for each group of rows across all columns. Alternatively, the user can choose to show only the first N items of the aggregated groups using the degree of interest function. 
In Figure~\ref{fig:aids}, for instance, only the first 10 African countries with a low and medium \textit{human development index} are displayed.

\subsection{Visualizing Matrices}
Although many tools offer support for time-series data (e.g., by showing sparklines), these tools usually do not support general matrices. % provided functionality does not cover operations necessary for general matrices. 
For example, the option to reorder the data points is usually missing, because it is not necessary for the time-series data. In our technique, adding a matrix to a table visualization introduces a second key for the columns of the matrix. We allow grouping of matrix columns based on this key. The individual groups of columns are then treated as separate matrices---they can be manually reordered, aggregated, and sorted, and the visual encoding of each group can be adjusted individually. For example, the years in the \textit{new HIV infections per 1,000 people} matrix and the \textit{AIDS-related deaths per 1,000 people} matrix in Figure~\ref{fig:aids} (e) introduce \textit{years} as the second key, which is then used to group these matrices by \textit{decades}. Here, the 2010s use a different visual encoding for the groups.

%This better illustrates the correlation between the two matrices. Thus, it can immediately be seen that an outburst of \textit{new HIV infections} in the 1990s in southern African countries (in particular Lesotho, Swaziland, Zimbabwe, Botswana, and South Africa) resulted in high \textit{AIDS-related death} rates about a decade later in the 2000s. From the box plot representation of the following decade it can also be seen that by the 2010s, the rates of \textit{new infections} and \textit{AIDS-related deaths} had evened out.

%In Figure \ref{fig:aids}, for example, we aggregated the the North American countries grouped by \textit{Human development index} to compare the distribution of number of \textit{People knowing they have HIV}. The histogram visualizations shows us, that while the in the countries with very high HDI have relatively low numbers of people knowing they have HIV, countries with high HDI contain higher numbers of of people knowing they have HIV than countries with medium HDI.

\begin{figure}[tb]
\centering
\includegraphics[width=1\linewidth]{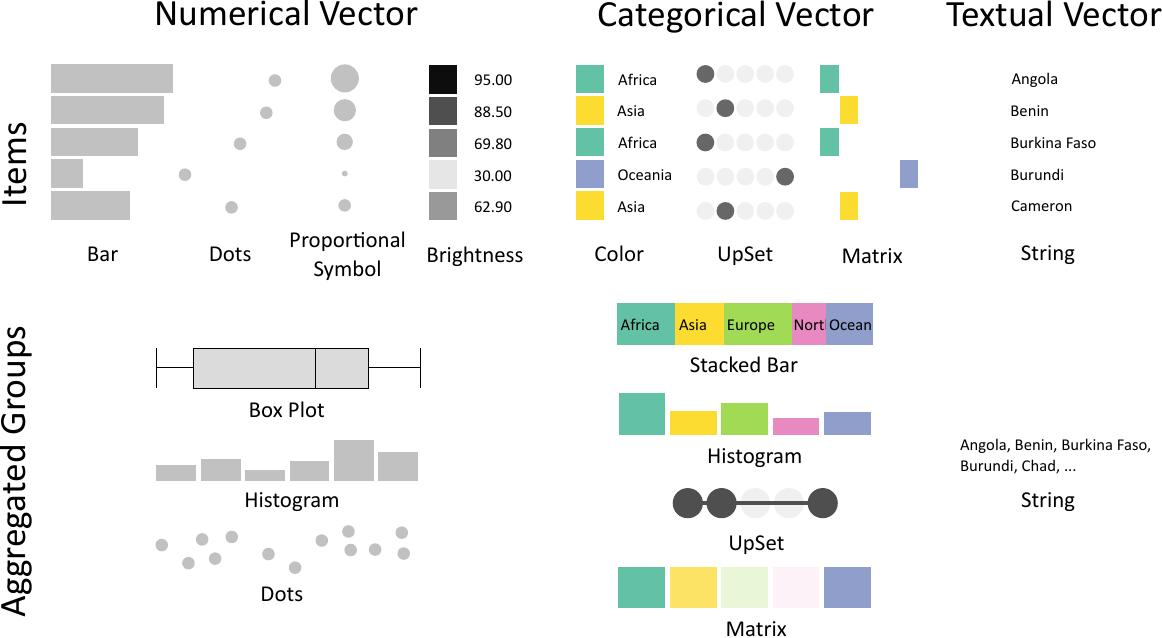}
\caption{Attribute column visualization techniques for items and aggregated groups by data type. \textbf{Numerical items} can be encoded with bars, dot plots, proportional symbols, or brightness. For \textbf{categorical items}, we offer color encoding plus labels, and two variants of matrix representations, one with and one without color used redundantly. All items can also be displayed as strings. \textbf{Numerical attributes} can be aggregated into box plots and histograms. \textbf{Distributions of categorical values} can be shown as a histogram, a stacked bar, a binary presence/absence matrix inspired by UpSet~\cite{lex_upset:_2014}, or an aggregated matrix with brightness encoding the frequency of individual categories in the group. An \textbf{aggregated textual attribute} shows examples of the group members.
%Aggregated textual vector indicates the content of the aggregated group---the category or intersection of categories forming the group, number of items in the group, and examples of the group members
}
\label{fig:visualizations_vector}
\end{figure}

\subsection{Encoding and Multiform Visualizations}

\begin{figure}[htb]
\centering
\includegraphics[width=1\linewidth]{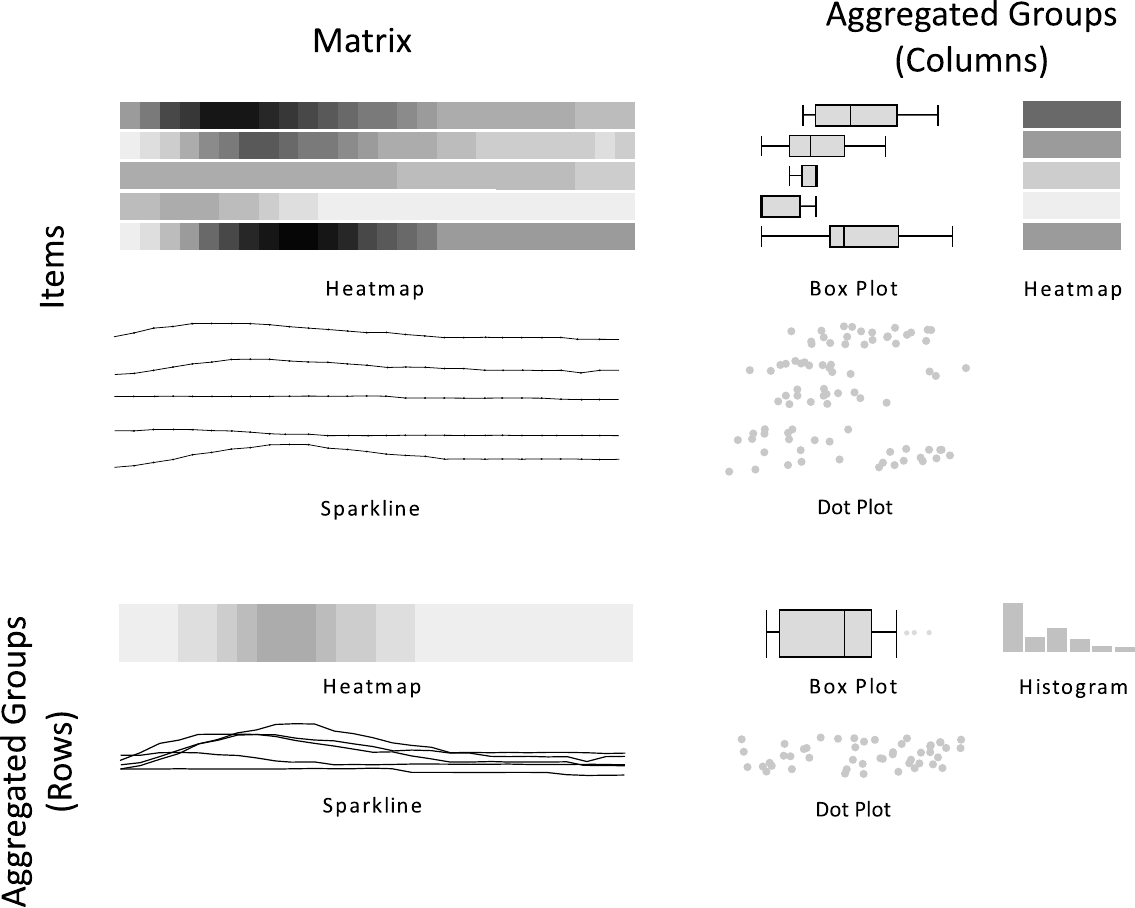}
\caption{Matrix visualization techniques. Matrix items can be encoded using brightness and as sparklines. Matrices can be aggregated in both column and row directions. When a matrix is aggregated in the column direction, a group of matrix columns within one row is merged into a single cell. The aggregated values can then be visualized using \new{box plots, dot plots, and heatmap}. When a matrix is aggregated in the row direction, a group of rows is merged into one row. Values of aggregated rows can be displayed using a heatmap and superimposed sparklines. A matrix aggregated in both directions is encoded using a box plot, histogram, or dot plot of all matrix values.}
\label{fig:visualizations_matrix}
\end{figure}

The table view encodes each selected column or matrix using one of multiple alternative visual encodings suitable for the data type, including bars, dots, proportional symbols, or brightness for numerical data; color or positional/matrix encoding for categorical data; and heatmaps for matrices. 

Following the multiform principle~\cite{lex_visbricks:_2011}, the visual encoding for each column can be changed on demand. For example, the default bar encoding a single numerical attribute can be interactively changed to a proportional symbol, if desired. Dedicated visual encodings are used for aggregates: box plots and histograms show the distribution of numerical values; stacked bars and histograms show relative frequencies of categories in an aggregate. A list of textual items is represented as a truncated list of examples. 
Figure~\ref{fig:visualizations_vector} gives an overview of the visual encodings available for numerical, categorical, and textual attributes with and without aggregation. Figure~\ref{fig:visualizations_matrix} summarizes how a matrix can be aggregated in column and row directions. \new{In theory, the aggregation choices for the matrix rows and columns should be symmetric. Our limited choices of visualizations for aggregated rows and columns stem from our design decision to show the aggregated rows with fixed height, whereas for aggregated columns the width is flexible and by default reflects the width of the matrix. Thus, most of the visualizations available for aggregated columns (e.g., box plots or dot plots) are not suitable for aggregated rows, as there would not be sufficient space.} 

We limit ourselves to these choices because they either offer \new{good perceptual properties (e.g., encoding by position)} or are very compact, thus allowing users to choose between perceptual accuracy and space utilization. We deliberately do not offer visual encodings that we consider to be problematic. For example, a bar representing an average of a group does not communicate any variability and is therefore not a suitable visualization for an aggregated attribute~\cite{streit_points_2014}. 

We chose a dash to encode missing values (Figure \ref{fig:fertility} (d)). We also considered a dedicated color, but dashes have the advantage that their visual saliency is lower (i.e., they do not draw as much attention), but are still clearly visible at all levels of detail.

\subsubsection*{Compact Encodings}

%\begin{figure}[htb]
%\centering
%\includegraphics[width=0.98\linewidth]{figs/figure6.pdf}
%\caption{Example of encodings at different scales. In the first column, the items are displayed at full height with white space separating the rows. If a textual label is part of the visualization, it is displayed at a readable size. Compact representations (columns two and three) remove white space and string labels. Some visualizations, such as the box plot or the categorical matrix, are simplified to account for the limited space.}
%\label{fig:visual_compression}
%\vspace{-7pt}
%\end{figure}

\begin{figure}[tb]
\centering
\includegraphics[width=\linewidth]{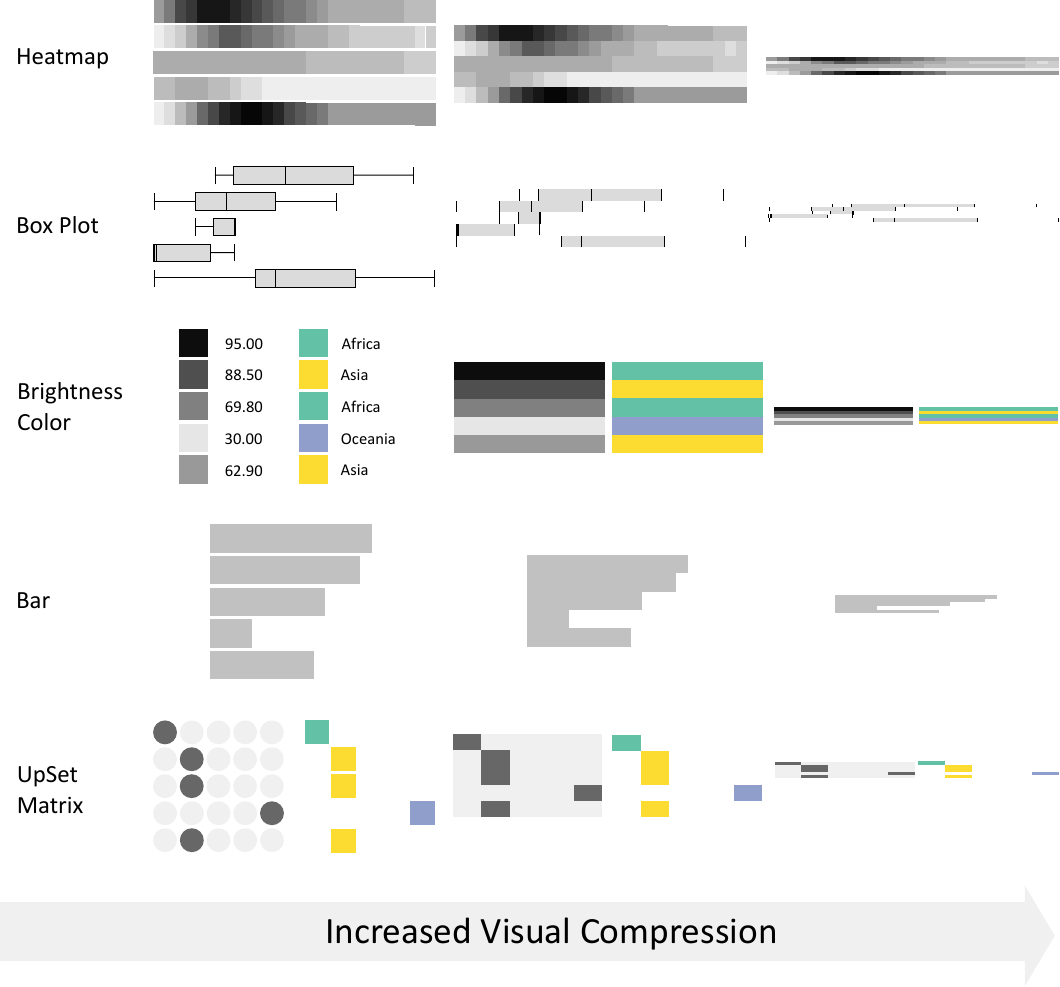}
\caption{Example of encodings at different scales. In the first column, the items are displayed at full height with white space separating the rows. If a textual label is part of the visualization, it is displayed at a readable size. Compact representations (columns two and three) remove white space and string labels. Some visualizations, such as the box plot, UpSet, or the matrix
representation, are simplified to account for the limited space.}
\label{fig:visual_compression}
\vspace{5mm}
\end{figure}

When the height of rows is reduced in overview mode, we take various measures to adapt the visualization to the diminished space.
%, as illustrated in Figure \ref{fig:visual_compression}. 
We not only make the visual representations smaller, but also reduce details and/or adapt the visualization. In the compact representation of box plots, for instance, we fill the available vertical space at the position of the box and indicate the start and end of the whiskers by drawing vertical tick marks. However, some visualizations, such as strings and proportional symbols, do not have an adequate downscaled version. We do not render such cells in overview mode. Examples of visual compression for individual visualization options can be found in Figure~\ref{fig:visual_compression}.

\subsection{Animated Transitions}
We support users in understanding changes in the visualization by applying animated transitions~\cite{heer_animated_2007}, as demonstrated in the accompanying video. Our implementation incorporates smooth transitions for the switch between overview and detail as well as for changes resulting from filter, sort, and aggregation operations. 

Instead of simply morphing item position, we apply staged transitions, where animations are split into multiple phases~\cite{heer_animated_2007}. In the first phase of a filter animation, for instance, we fade out the filtered rows and then move up the remaining rows of the table to fill the white space. This animation is designed to help users understand why rows outside the viewport become visible at the bottom of the table. Similarly, when items in a group are collapsed, we first fade out the items and then gradually reduce the height of the empty group area to the fixed height of the aggregated group.

\subsection{Combining Columns}

\new{Giving users the ability to flexibly combine columns supports various tasks. Users} can interactively create combined columns by dragging either existing ones on an empty container or one column onto another. The possible combinations are specific to the data type of the column.

\begin{figure}[tb]
\centering

        \includegraphics[width=1.05\columnwidth]{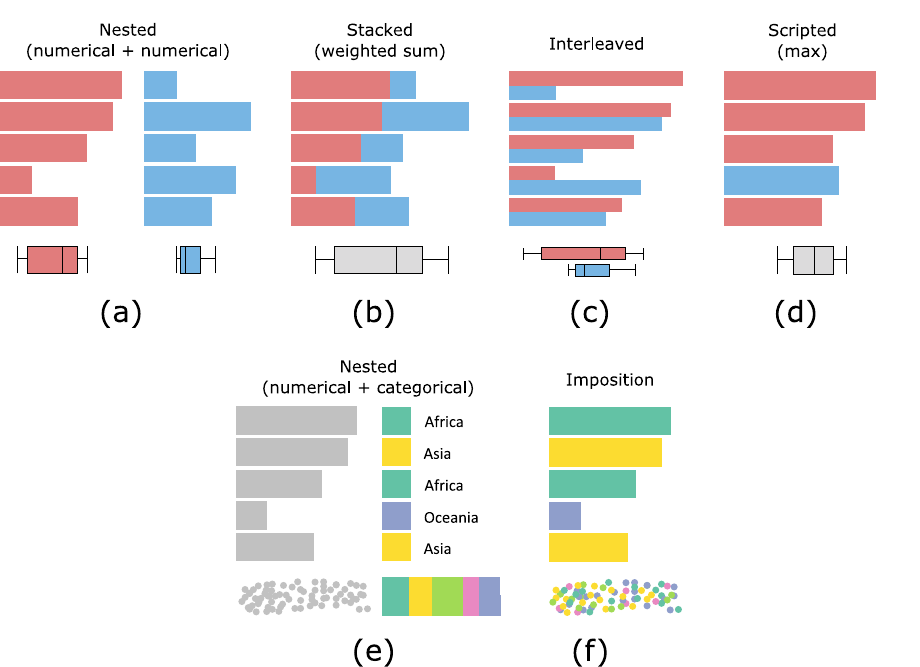}
    \caption{Possible column combinations: (a, e) nested column that semantically groups columns of various types; (b) stacked column that creates a stacked bar plot based on multiple weighted numerical columns; (c) interleaved column that stacks the visualizations of multiple numerical columns; (d) scripted column that, in this case, visualizes only the maximum values of selected columns; and (f) column imposition where the marks of a numerical column are colored by the imposed categorical column.}
    \label{fig:combining}
\end{figure}

\begin{figure*}[tb]
\includegraphics[width=0.98\linewidth]{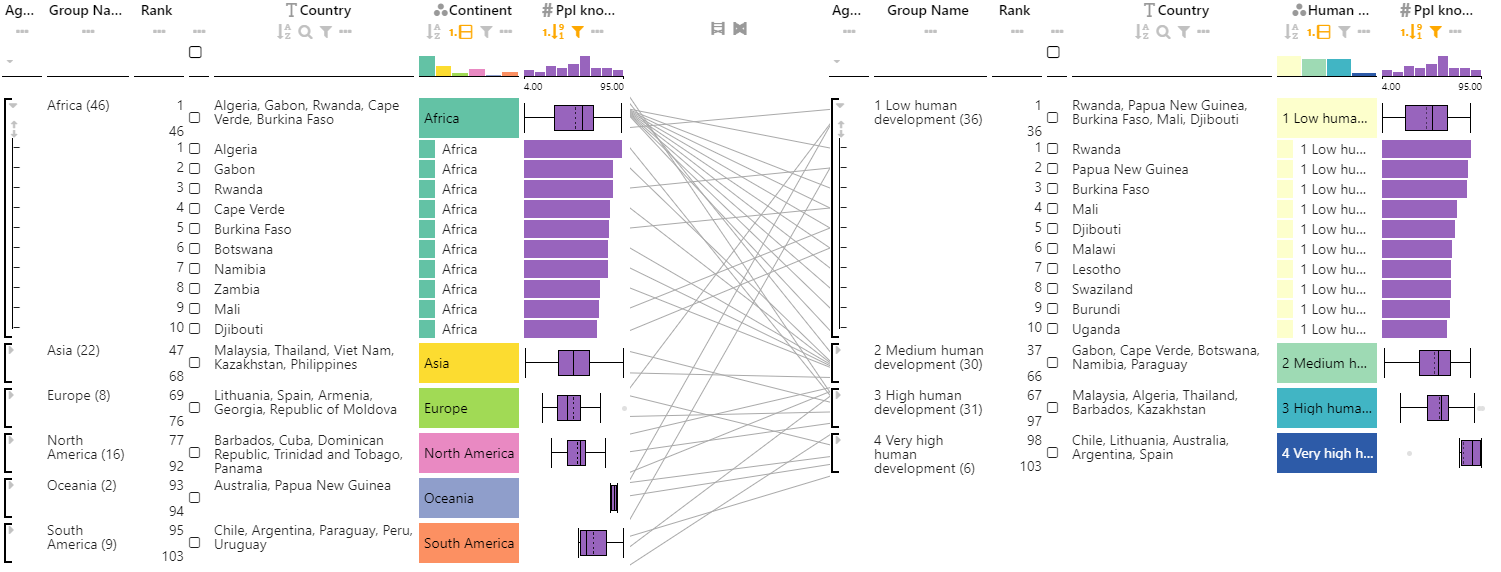}
\caption{Comparison of two table segments. The segment on the left shows countries grouped by \textit{continent}. The segment on the right visualizes countries grouped by the \textit{human development index}. Both table segments are ranked by the number of \textit{people knowing they have HIV}. The steeper the angle of the lines connecting the two instances, the greater is the change in the ranking. Bands show relationships between aggregated groups.}
\label{fig:slope_graphs}
\end{figure*}

The most basic combined column is a \textbf{nested column}, shown in Figure~\ref{fig:combining} (a) and (e). It encloses multiple individual columns by adding a joint header above all columns contained. Nested multiple columns are useful for creating semantic groups. The nested column is the most flexible column combiner that works for all types and can mix columns of different types.

Taggle also enables users to create \textbf{stacked columns}~\cite{gratzl_lineup:_2013, carenini_valuecharts:_2004} by combining two or multiple numerical columns to create a weighted sum of the items and where the individual contributions are represented as \textit{stacked bars} (see Figure~\ref{fig:combining} (b)). Users can interactively change the weights of individual columns by adapting their widths. Stacked columns can be used to create a ``score'', which, in turn, can be used to create rankings. Aggregate representations for stacked columns are shown as box plots, where the values feeding the box plots are the weighted sums of the composing values.

To enable a more effective comparison of items across multiple columns, an \textbf{interleaved column} (Figure~\ref{fig:combining} (c)) stacks the encoded values from multiple numerical columns vertically. Depending on whether the row is an item or group, the stacked representations can be made from bars or dots, or, in case an aggregate is interleaved, from a box plot. 

With \textbf{imposition columns}, users can color the visual marks (bar, proportional symbol, etc.) of a numerical column by the color coding of a categorical attribute, as shown in Figure~\ref{fig:combining} (f).

Taggle also enables more complex combinations, based on a set of predefined functions, such as minimum, maximum (Figure~\ref{fig:combining} (d)), and mean, for combining multiple numerical attributes into a single numerical column. In addition, users can add \textbf{scripted columns} that allow them to define their own functions via a scripting interface~\cite{gratzl_lineup:_2013}.

\subsection{Sorting and Grouping of Column Subsets}

Although Taggle focuses primarily on tabular visualization, keeping items in constant rows across all columns, it also supports splitting a table into multiple segments and sorting and grouping each instance independently. %\new{This way users can observe the relationship between individual table segments, as done in the list view from Ploceus et al.~\cite{liu_network-based_2011, liu_ploceus:_2014}.} 
To encode the relationships between table segments, we utilize slope graphs for connecting individual items of the tables compared~\cite{tufte_visual_2001}$^{, p. 156}$ or bands for showing relationships between aggregated groups~\cite{lex_stratomex:_2012a, gratzl_domino:_2014}, de facto enabling users to create hybrid tabular/overview representations (see Figure~\ref{fig:slope_graphs}), and in the extreme, even visualization techniques such as parallel sets~\cite{kosara_parallel_2006}.

%%% END OF CONCEPT

%%%%%%%%%%%%%%%%%%%%%%%%%%%%%%%%%%%%%%%%%%%%%%%%%%%%%%%%%%%%%%%%

\section{Implementation}

In the demo application\footnote{\url{https://taggle.caleydoapp.org/}}, users can switch between multiple preloaded datasets, upload datasets, and download existing datasets in various formats. Users can locally save and restore a Taggle table together with the analysis session that includes the history of all user interactions. 

The Taggle feature set is fully integrated into the \textit{LineUp.js} library\footnote{\url{https://lineup.js.org/}}, which is written in TypeScript and available as open source\footnote{\url{https://github.com/lineupjs/lineupjs/}}. A demo version can be accessed at \url{https://taggle.caleydoapp.org/}. Making Taggle available as an open-source library increases the potential for adoption of the technique. Taggle is also designed to be combined with other techniques. To that end, we provide various interfaces. For example, the library can be embedded in Jupyter Notebooks\footnote[4]{\url{https://jupyter.org/}}\textsuperscript{,}\footnote{\url{https://github.com/lineupjs/lineup_widget/}} and used as an HTML widget\footnote{\url{https://www.htmlwidgets.org/}}\textsuperscript{,}\footnote{\url{https://github.com/lineupjs/lineup_htmlwidget}}, which allows integration into Shiny applications\footnote{\url{http://shiny.rstudio.com/}}, R~Notebooks\footnote{\url{http://rmarkdown.rstudio.com/r_notebooks.html}}, Anuglar.js\footnote{\url{https://angularjs.org/}}, Vue.js\footnote{\url{https://vuejs.org/}}, and React.js\footnote{\url{https://reactjs.org/}}. We provide examples for how to embed Taggle in each of these frameworks in the repository. 
Note that Taggle can also be embedded as a component inside a larger web-based application. The case study described in the following section is based on the Ordino visual cancer analysis tool~\cite{streit_ordino:_2019}. The server-side of Ordino retrieves over 500 GB of cancer data from a PostgreSQL database. Complex aggregation queries that need to iterate over a large set of table entries are handled by the database, while the client-side with Taggle then receives only the data subset needed for rendering. 

%%%%%%%%%%%%%%%%%%%%%%%%%%%%%%%%%%%%%%%%%%%%%%%%%%%%%%%%%%%%%%%%

%\section{Evaluation}

%We evaluate Taggle by means of two methods that are both widely used to validate visualization systems~\cite{sedlmair_design_2012, lam_empirical_2012}: a qualitative user study, and a case study conducted by a domain expert on complex genomics data for the purpose of drug target discovery. 

\section{Case Study: Drug Target Discovery}
\label{ssec:cs-ordino}
\new{Taggle was developed in tight collaboration with domain experts working on a drug discovery team at a pharmaceutical company.}
We \new{demonstrate} Taggle by means of a case study conducted on complex genomics data for the purpose of drug target discovery. The case study summarizes an analysis session carried out by \new{one of our collaborators}. For the case study, we integrated Taggle into the Ordino Target Discovery Platform~\cite{streit_ordino:_2019} that provides access to the required cancer genomics data\footnote{\url{https://ordino.caleydoapp.org/}}. Note that the collaborator has experience using interactive visualization tools and was involved in all phases of the project and provided continuous feedback during the development. For the case study, the domain expert operated Taggle himself without the help of visualization experts.

In order to identify potential drug targets in a set of tumor types, the analyst performs experiments with cancer cell lines---cultured cells that are derived from tumors and that can proliferate indefinitely in the laboratory. These cell lines are characterized by various properties, such as tumor type (lung cancer, prostate cancer, etc.) and the set of genes that are mutated. 

One very important gene in the context of cancer is \textit{TP53}. It encodes the p53 protein, whose presence is known to suppress the uncontrolled division of cells. However, when \textit{TP53} is mutated---which is the case for over 50\% of cancer patients---it can lose its suppressing function, which results in tumor growth. Due to its important role, scientists want to know whether \textit{TP53} is mutated in a set of cell lines. However, the mutation status of \textit{TP53} is not always known. It has recently been shown that the mean expression level (expression is a measure of the activity of genes) of 13 genes that are biologically related to \textit{TP53} is correlated with its mutation status. The expression level of these genes can hence be used to predict the mutation status of \textit{TP53}~\cite{jeay_distinct_2015}. 
%to make the case study not too complex, we decided to remove the MDM2 sensitivity aspect for now}
%Furthermore, cell lines that are sensitive to the loss of the gene \textit{MDM2}, an important negative regulator of \textit{TP53}, are expected to not harbour a \textit{TP53} mutation. Because, if \textit{TP53} is intact but proper functioning is prevented by \textit{MDM2}, the down-regulation of \textit{MDM2} allows \textit{TP53} to induce cell death. 

In this case study, the analyst first wants to find out how well this predictor works for the set of cell lines contained in the database. Based on this knowledge and other criteria, the analyst then wants to select cell lines for a wet-lab experiment.

The analyst starts by loading a list of 1,009 cell lines from the public CCLE dataset~\cite{barretina_cancer_2012} into Taggle. By default, the table contains a textual column representing the \textit{names} of cell lines and a categorical column indicating \textit{tumor type}. Since only a subset of tumor types is of interest, the analyst filters for \textit{astrocytoma/glioblastoma} (type of cancer of the brain), \textit{bone sarcoma}, \textit{melanoma}, and \textit{non-small-cell lung cancer (NSCLC)}, after which 255 cell lines remain.

\begin{figure}[t]
\centering
\includegraphics[width=1\linewidth]{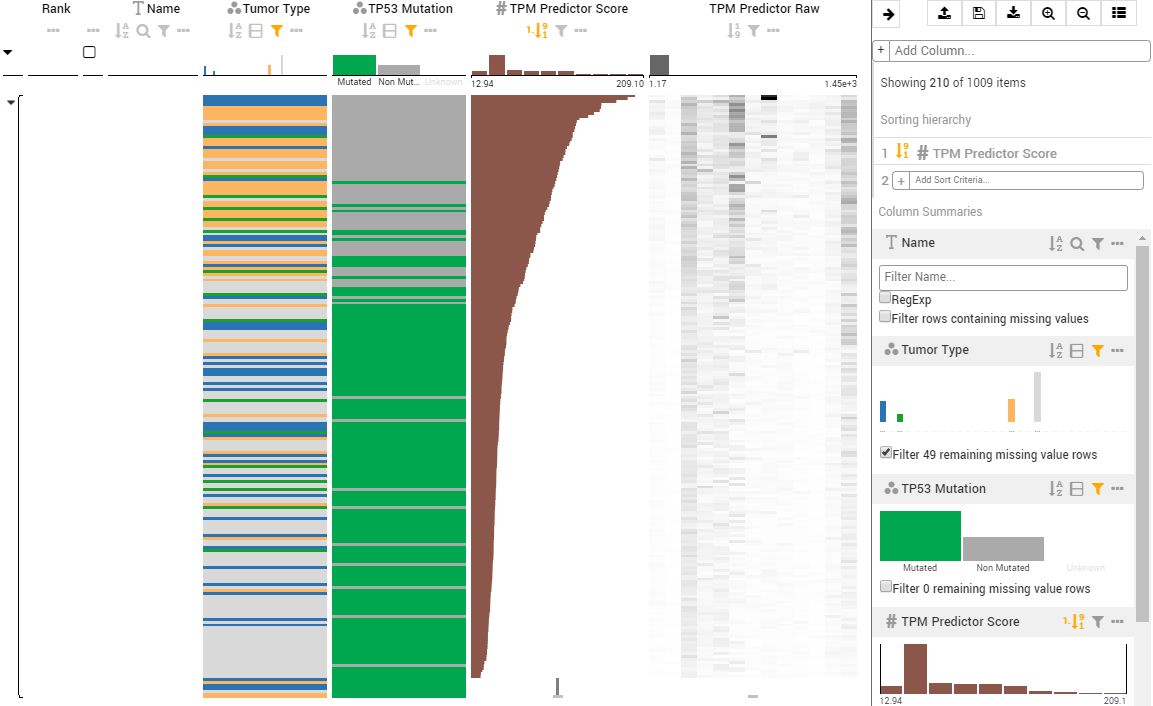}
%\vspace{-10pt}
\caption{After sorting the cell lines by the \textit{TP53 predictor score} (brown), the analyst notices that those with a low average score are much more likely to be mutated (green). From this observation, the analyst concludes that predicting the mutation status based on the average expression of the 13 genes that constitute the predictor score works reasonably well.
%\vspace{-10pt}
}  
\label{fig:case_study_TP53_predictor}
\end{figure}

As the analyst wants to investigate the \textit{TP53} gene, he loads a categorical column with the \textit{mutation status} (mutated vs.~nonmutated) and a textual column that provides further details about the mutation (if present). According to the mutation histogram in the data selection panel, the status is unknown for 59 cell lines. To investigate the effectiveness of the 13 genes in predicting the \textit{TP53} status, the analyst loads the average expression of these genes together with a matrix column containing the individual expression values. Furthermore, he hides cell lines with unknown mutation status. After sorting the table by average expression in descending order and switching to the overview (see Figure \ref{fig:case_study_TP53_predictor}), the analyst observes the overall good correlation between expression and mutation status: there is a clear enrichment of \textit{TP53} mutants among the cell lines with a low score. 

In order to test whether the correlation is present for all selected tumor types, the analyst groups the table by tumor type. He observes that the prediction seems to work particularly well for the \textit{astrocytoma/glioblastoma} cell lines (almost perfect separation between mutated and nonmutated) and further investigates this observation by also stratifying by mutation status and aggregating all groups (see Figure~\ref{fig:case_study_aggregated_group_comparison}). The expression box plots show good separation for \textit{astrocytoma/glioblastoma} and \textit{melanoma}, whereas the expression ranges are overlapping for \textit{NSCLC}.

Having confirmed that the prediction of the \textit{TP53} mutation status works reasonably well in several tumor types, the analyst wants to select a set of cell lines for a wet-lab experiment. He is interested in \textit{melanoma} cell lines that have no \textit{TP53} mutation. Furthermore, the activity of \textit{CDKN2A}, another important tumor suppressor gene, should be impaired due to a reduced number of \textit{CDKN2A} gene copies in the genome. The analyst removes the mutation status grouping, includes cell lines for which it is unclear whether \textit{TP53} is mutated, and unfolds the \textit{melanoma} cell lines group. Based on the ranking, he decides to consider all cell lines with unknown \textit{TP53} mutation status and a \textit{TP53} predictor score greater than 110 as nonmutated. He adds a column with the \textit{CDKN2A} relative copy number, sorts by it in ascending order, and filters out missing data. Finally, he selects the top hits of the resulting list (see Figure~\ref{fig:case_study_melanoma_drilldown}). All these cell lines fulfill the analyst's requirements.

%Was there any insight that was not anticipated? What worked well and less well?
\subsubsection*{Expert Feedback}
Our collaborators initially planted the seed that led to the development of Taggle by pointing out restrictions they face in current drug discovery tools. They particularly mentioned the need of seamlessly combining overview and details in tabular data analysis for drug discovery.

During the conception and development of Taggle, we had biweekly feedback sessions and in-depth discussions with our collaborators on every aspect of both the concept and the visual interface. The most critical feedback on early prototypes was about the limited rendering performance that hindered their use in real-world scenarios. After making the prototypes more scalable, we received valuable and very detailed feedback on a conceptual level but also regarding the usability of the prototype implementation. For example, the user interface workflow and visual encoding of the hierarchical grouping and sorting capabilities led to confusion. 
%This was also confirmed by multiple user study participants (see Section~\ref{ssec:user-study}). 
We resolved this problem by introducing an explicit sorting and grouping hierarchy in the data selection panel (see Sections~\nameref{ssec:sorting} and ~\nameref{ssec:grouping}). Based on follow-up feedback, we also added the capability of controlling the order of groups, to sort them by number of items or by group name, for instance. 

The fact that Taggle recently replaced the LineUp technique as a core component in the Ordino drug discovery tool~\cite{streit_ordino:_2019}, which is in productive use at Boehringer Ingelheim, demonstrates that the domain experts are convinced of its effectiveness and added value.

In additional high-level feedback, the domain experts mentioned that they would like to confirm the statistical significance of visual patterns they see in the overview as well as between groups of items. \new{However, this approach could easily lead to incorrect inferences, unless some precautions are taken~\cite{dwork_reusable_2015, zgraggen_investigating_2018}}. In a follow-up project, we are working on a solution that supports such confirmatory analysis in a way that users can understand without being trained in statistics~\cite{eckelt_tourdino:_2019}.

%%%%%%%%%%%%%%%%%%%%%%%%%%%%%%%%%%%%%%%%%%%%%%%%%%%%%%%%%%%%%%%%
\section{Discussion and Limitations}

\begin{figure*}[htb]
\centering

\includegraphics[width=1\linewidth]{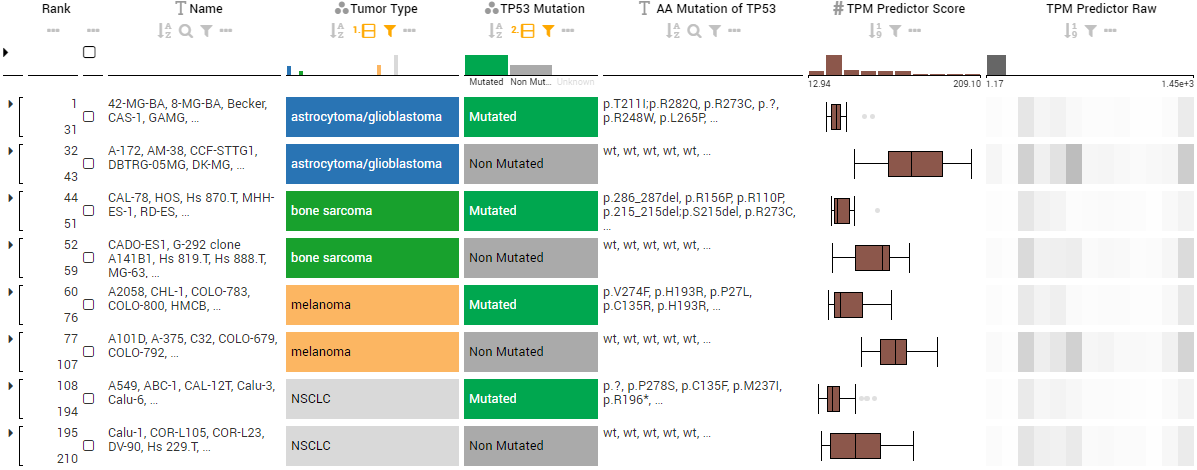}
\caption{The analyst groups the cell lines first by the attribute \textit{tumor type} and then by \textit{TP53 mutation status}. For the tumor type \textit{astrocytoma/glioblastoma}, the box plots representing the \textit{TP53 predictor score} show a clear separation between the groups \textit{mutated} and \textit{nonmutated}. For the other tumor types, the whiskers of the box plots overlap, indicating that the predictor score does not work as effectively.}
\label{fig:case_study_aggregated_group_comparison}
\end{figure*}

\begin{figure*}[htb]
\vspace{2mm}
\centering
\includegraphics[width=1\linewidth]{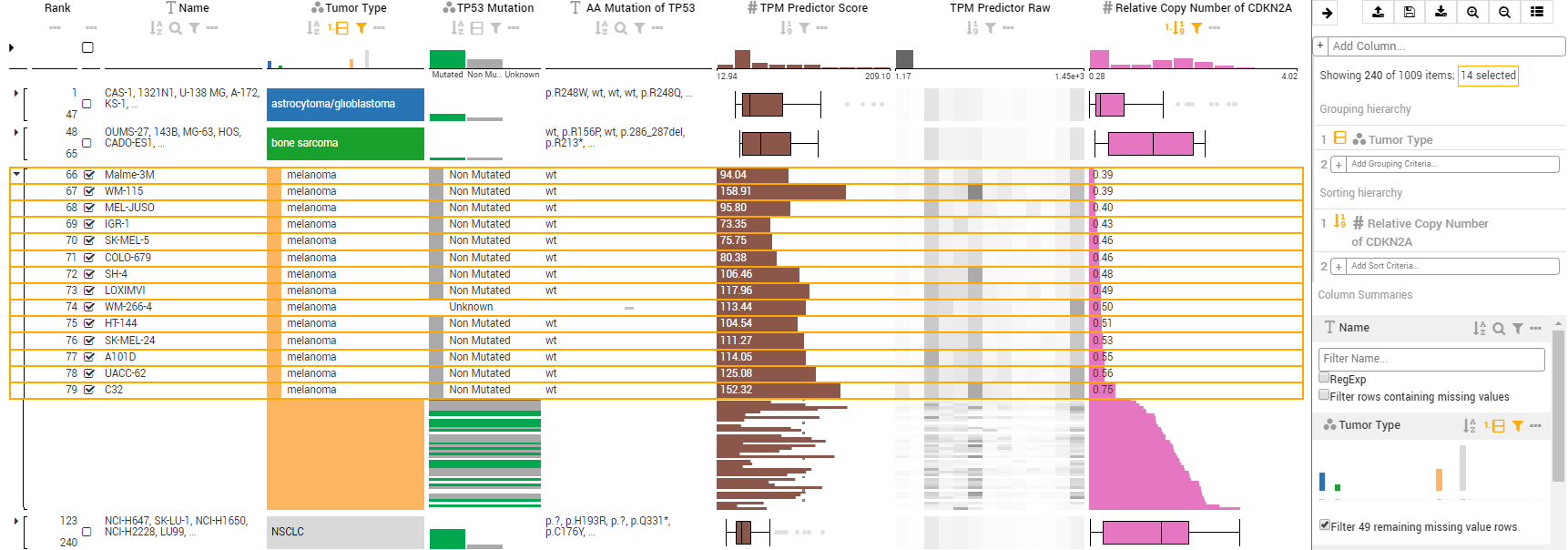}
\caption{Continuing from the visualization state shown in Figure~\ref{fig:case_study_aggregated_group_comparison}, the analyst removes the grouping on the \textit{TP53 mutation} column and unaggregates the \textit{melanoma} group to inspect the cell lines in further detail. With the goal to find cell lines for a wet-lab experiments, the analyst adds the \textit{copy number value} of \textit{CDKN2A} as an additional column (shown in pink). Finally, he selects cell lines that have a low \textit{copy number value} and are either nonmutated or have unknown mutation status and a \textit{TP53 predictor} score above 110.
}
\label{fig:case_study_melanoma_drilldown}
\end{figure*}

Revisiting our discussion of visualization techniques for tabular data (overview, projection, tabular, and MCV techniques), we argue that Taggle is primarily a tabular visualization technique, as it retains a tabular layout and encodes data within a cell, but also has some aspects of an overview technique due to its capabilities to aggregate and its ability to sort and group subsets of columns independently. Interactive definition of groups and their aggregation in summary visualizations, such as box plots and histograms, provides a meaningful overview even for large datasets and enables an intuitive comparison of grouped data subsets (Figure~\ref{fig:case_study_aggregated_group_comparison}). At the same time, Taggle enables the exploration of items at a detailed level to identify their precise properties (Figure~\ref{fig:case_study_melanoma_drilldown}). We also designed Taggle so that it can be used within an MCV framework. 

This combination sets Taggle apart from existing tabular techniques, which provide only a coarse overview of items (e.g., using the lens technique, which is insufficient for representation or comparison of large datasets) or lack interactivity, which is essential to the exploration process.

\subsection{Scalability}

Taggle scales to more than 1 million rows on a modern browser, as demonstrated when loading the \textit{1M Random Dataset} in the demo application. We achieve this performance by leveraging rendering optimizations, which ensure that only visible rows are processed. Although the rendering time stays almost constant, larger datasets require more time for data operations, such as sorting, grouping, or computing histograms---which always need to be done for the full dataset. The performance depends on the number of CPU cores available on the client machine, as the workload is distributed between multiple parallel web workers, if possible.

To demonstrate the computational scalability, we executed performance measurements for common operations on five datasets consisting of 100, 1,000, 10,000, 100,000 and 1,000,000 data items. Each dataset consisted of one textual, two numerical, and two categorical attributes generated with uniform distribution. For each tested operation, we measured the time between triggering the operation (e.g., pressing the sort button) and the end result appearing on screen. Animation is not useful when rearranging large datasets; hence, it is disabled by default for such datasets. To make the results comparable across all datasets, we disabled animations for all conditions when benchmarking. For measurements we used the performance profiler from Google Chrome DevTools (v.\ 71.0.3578.98). We repeated each measurement five times. Table~\ref{table:performance} presents the average times in milliseconds. The tests were done on a machine with Intel Core i7-5930K processor (3.5 GHz, 6 cores), 32 GB RAM, NVIDIA GeForce GTX 970 graphics card. Note that the browser-based tracking tool may slightly decrease the actual performance.

\begin{table}[]
\vspace{2mm}
\footnotesize
\begin{tabular}{r l r r r r r}
%\rowcolor[HTML]{E3DBD3} 
                    &           & \textbf{100}    & \textbf{1K}     & \textbf{10K}   & \textbf{100K}   & \textbf{1M}     \\ 
                            \cmidrule{3-7}
%\rowcolor[HTML]{f0f0f0}
\textbf{Load }          &            & 529  & 545    & 611  & 1,012 & 4,107 \\ 
\addlinespace
\multirow{2}{*}{\textbf{Sort numerical}} & DM                  & 321  & 338  & 358  & 643  & 2,642   \\ 
%\rowcolor[HTML]{f0f0f0} 
 & OM            & 288  & 681  & 741  & 970  & 3,626 \\ 
\addlinespace
\multirow{2}{*}{\textbf{Sort grouped}} & DM            & 324  & 324  & 381  & 518  & 1,911   \\ 
%\rowcolor[HTML]{f0f0f0} 
 & OM      & 306  & 661  & 743    & 923  & 2,830 \\ 
\addlinespace
\multirow{2}{*}{\textbf{Sort textual}} & DM                & 300  & 367    & 397  & 728  & 3,639   \\ 
%\rowcolor[HTML]{f0f0f0} 
&OM        & 302  & 647  & 730  & 1,069 & 5,075 \\ 
\addlinespace
\multirow{2}{*}{\textbf{Filter numerical}} & DM                  & 407  & 415  & 460  & 598  & 1,442 \\ 
%\rowcolor[HTML]{f0f0f0} 
 & OM          & 419    & 1,883 & 1,745 & 1,876 & 2,858 \\ 
\addlinespace
\multirow{2}{*}{\textbf{Filter categorical}} & DM      & 357  & 435    & 475    & 562  & 1,372 \\ 
%\rowcolor[HTML]{f0f0f0} 
 & OM & 403    & 1,982 & 1,079 & 1,196 & 2,261 \\ 
%Switch to Overview Mode                    & 192  & 507  & 536  & 628    & 1,838 \\ 
%\rowcolor[HTML]{f0f0f0} 
%Painting                       & 6    & 155  & 173    & 162  & 160  \\ \hline
\end{tabular}
\vspace{5mm}
\caption{
Completion time in milliseconds for various operations using five datasets with 100 to 1 million items. DM indicates operations performed in detailed mode, and OM indicates operations performed in overview mode. 
}
\vspace{-12mm}
\label{table:performance}
\end{table}

Since the full dataset needs to be loaded into memory first, the size of the data table determines the loading time. Naturally, the number of rendered items also influences the run-time performance. Although we optimize the rendering to process only visible rows, there can be notable performance differences between detail and overview mode, since the number of rendered items is much larger in overview mode. For example, in our full-HD setup with viewport size 1,387$\times$882 pixel, detail mode (DM) allowed for 39 item rows, but in overview mode (OM) we had up to 775 rows on screen. Note that due to the design decision that every item is at least one pixel high, the table grows out of the visible screen space for larger tables. Table~\ref{table:performance} shows that for the smallest dataset, where the number of rendered elements is low in either case, the performance difference between overview and detail mode is minimal. For other datasets, the time necessary for preparing and rendering the elements is much more apparent.

\subsection{Aggregation of Categorical and Textual Attributes}

Although there are numerous possibilities for aggregation for numerical data---ranging from aggregation in data space (mean, median) to spatial aggregation (box plots, histograms)---the options are limited for categorical and textual data. In our prototype we offer three possibilities for aggregation of categorical data: a matrix, a histogram, and a distribution bar. Due to spatial restraints and the limited scalability of the color channel, there are limits with respect to the number of categories that can be sensibly encoded this way. Taggle uses a predefined color scheme with 22 distinct colors. However, if a categorical attribute has more than 22 categories, we treat the column as textual. Colors are automatically assigned, but users also have the possibility to adjust colors manually to resolve cases where colors are repeated between columns.

%We address this problem by treating categorical attributes with more than 22 categories as textual attributes.

%\subsection{Use of Color}
%Choosing appropriate colors is an essential part of data visualization. We automatically assign colors to numerical columns and categories within categorical columns from predefined D3 color schemes\footnote{\url{https://github.com/d3/d3-scale-chromatic/}}. However, picking colors for encoding categorical data is problematic in general because of the limited number of distinguishable colors. Taggle uses a color scheme with 22 distinct colors. If a categorical attributes has more than 22 categories, we treat the column as textual. It may also happen that the automatically selected colors are repeated between columns. While there is no ideal solution to this problem, users have the possibility to assign colors manually.

Aggregation of textual attributes, however, is even more limited. In our prototype implementation, we list a sample of items from the aggregated group to summarize the group's content using the order of the items. An alternative approach would be to select samples based on other criteria such as frequency of occurrence. This approach, however, is practical only for data attributes with repetitive values. %If each item has a unique value of the attribute (e.g., name of the country in the AIDS dataset), this approach is not feasible.

\subsection{Automatic Aggregation}

In the design process, we investigated methods for automatically aggregating rows and columns, with the goal of increasing scalability. For example, when in overview mode, we tried to automatically aggregate groups to make space for user-selected rows that are shown with increased height. We found, however, that users had difficulties understanding the unexpected changes and subsequently interpreting the individual items and aggregated groups. As this violated the \textit{discriminability} design guideline proposed by Elmqvist and Fekete~\cite{elmqvist_hierarchical_2010}, we removed the automatic aggregation feature. %This may be due to perceptional phenomena, such as change blindness, or due to the fact that it can be hard to understand changes that the user did not actively trigger. 
Instead, as part of future work, we plan to implement and evaluate a recommendation approach that suggests possible layout changes without automatically applying them.

\new{
\subsection{Stacking of Matrices and Vectors}
Our current prototype supports grouping of matrix columns based on a categorical attribute (see Figure~\ref{fig:aids}), but provides no means of sorting and filtering the matrix columns. Furthermore, it is not possible to stack additional attributes on top of a matrix, as done, for instance, in Figures 4 and 6 presented in \cite{cherniack_integrated_2017}. However, we plan to address this technical limitation in future versions.}

%%%%%%%%%%%%%%%%%%%%%%%%%%%%%%%%%%%%%%%%%%%%%%%%%%%%%%%%%%%%%%%%

\section{Conclusion and Future Work}

In this work, we presented Taggle, an item-centric, spreadsheet-like visualization technique for exploring and presenting large and complex tables. Taggle is unique among tabular data visualization techniques due to its ability to dynamically aggregate subsets of a table, which allows users to flexibly drill-down into details of large tables while keeping the overview as context. 

The open-source implementation presented as part of this work goes beyond a research prototype, providing a rich set of visual encodings and rendering optimizations that make it scale to a million items. Taggle can be used as a standalone tool but also integrated as a widget into MCV systems or notebook-style environments such as R Markdown or Jupyter Notebooks.
%Taggle with traditional spreadsheet applications, combining the powerful logic and math enabled by spreadsheets, and the ease of exploration and discovering trends and outliers in Taggle. 

As part of future work, we plan to integrate data-driven guidance capabilities into Taggle, as implemented in StratomeX~\cite{streit_guided_2014}. Following the idea of guided visual exploration, we plan to assist users in finding correlated attributes or similar groups based on their input.

\begin{acks}
\new{We thank Bikram Kawan and Martin Ennemoser for their contributions to the initial prototype implementation as well as Christian Haslinger and Andreas Wernitznig for providing valuable conceptual feedback.}
\end{acks}

\begin{dci}
\new{Samuel Gratzl, Holger Stitz, Alexander Lex, and Marc Streit hold shares and/or are employed by datavisyn GmbH, which provides its customers with support for using and deploying the open source Taggle software.}
\end{dci}

\begin{funding}
\new{This work was supported in part by Boehringer Ingelheim Regional Center Vienna; the State of Upper Austria (FFG \#851460); the Austrian Science Fund (FWF P27975-NBL); and the National Science Foundation (NSF IIS 1751238).}
\end{funding}

\bibliographystyle{SageV}
\bibliography{taggle.bib}

\end{document}